\def\papertitle{A Reduced Multiple Gabor Frame for Local Time Adaptation of the Spectrogram}
\def\paperauthorA{Marco Liuni}
\def\paperauthorB{Axel R\"oebel}
\def\paperauthorC{Marco Romito}
\def\paperauthorD{Xavier Rodet}

\documentclass[twoside,a4paper]{article}
\usepackage{dafx_10}
\usepackage{amsmath,amssymb,amsfonts,amsthm}
\usepackage{euscript}
\usepackage[applemac]{inputenc}
\usepackage[T1]{fontenc}
\usepackage{ifpdf}

\usepackage[english]{babel}
\usepackage{caption}
\usepackage{subfig, color}

\usepackage{latexsym}
\usepackage{mathrsfs}
\usepackage{psfrag}

\ninept

\usepackage{times}

\newif\ifpdf
\ifx\pdfoutput\relax
\else
   \ifcase\pdfoutput
      \pdffalse
   \else
      \pdftrue
\fi

\ifpdf 
  \usepackage[pdftex,
    pdftitle={\papertitle},
    pdfauthor={\paperauthorA, \paperauthorB, \paperauthorC, \paperauthorD},
    colorlinks=false, 
    bookmarksnumbered, 
    pdfstartview=XYZ 
  ]{hyperref}
  \usepackage[pdftex]{graphicx}
  \usepackage[figure,table]{hypcap}
\else 
  \usepackage[dvips]{epsfig,graphicx}
  \usepackage[dvips,
    colorlinks=false, 
    bookmarksnumbered, 
    pdfstartview=XYZ 
  ]{hyperref}
  \usepackage[figure,table]{hypcap}
\fi

\title{\papertitle}

\fouraffiliations{
\paperauthorA, \sthanks{This work is supported by grants from Region Ile-de-France}}
{\href{http://www.math.unifi.it/}{Universit di Firenze, Dip. di Matematica U. Dini,} \\ Viale Morgagni, 67/a - 50134 Florence - ITALY\\ \href{http://anasynth.ircam.fr/}{IRCAM - CNRS STMS, Analysis/Synthesis Team,}\\1, place Igor-Stravinsky - 75004 Paris - FRANCE \\
{\tt \href{mailto: marco.liuni@ircam.fr}{marco.liuni@ircam.fr}}
}
{\paperauthorB,}
{\href{http://anasynth.ircam.fr/}{IRCAM - CNRS STMS, Analysis/Synthesis Team,}\\1, place Igor-Stravinsky - 75004 Paris - FRANCE \\
{\tt \href{mailto: axel.roebel@ircam.fr}{axel.roebel@ircam.fr}}
}
{\paperauthorC,}
{\href{http://www.math.unifi.it/}{Universit di Firenze, Dip. di Matematica U. Dini,} \\ Viale Morgagni, 67/a - 50134 Florence - ITALY \\ {\tt \href{mailto: marco.romito@math.unifi.it}{marco.romito@math.unifi.it}}
}
{\paperauthorD,}
{\href{http://anasynth.ircam.fr/}{IRCAM - CNRS STMS, Analysis/Synthesis Team,}\\1, place Igor-Stravinsky - 75004 Paris - FRANCE \\
{\tt \href{mailto: xavier.rodet@ircam.fr}{xavier.rodet@ircam.fr}}
}

\newcommand{\m}{\mathbb}

\newcommand{\ca}{\hat}
\newcommand{\de}{\mathrm{d}}

\newcommand{\ldue}{$L^2(\m{R})~$}
\newcommand{\mldue}{L^2(\m{R})}

\newcommand{\beq}{\begin{equation}}
\newcommand{\eeq}{\end{equation}}
\newcommand{\lsc}{\langle}
\newcommand{\rsc}{\rangle}

\begin{document}
\ifpdf 
  \DeclareGraphicsExtensions{.png,.jpg,.pdf}
\else  
  \DeclareGraphicsExtensions{.pdf}
\fi

\maketitle

\begin{abstract}
In this paper we propose a method for automatic local time adaptation of the spectrogram of an audio signal, based on its decomposition within a Gabor multi-frame. The sparsity of the analyses within each individual frame is evaluated through the Rnyi entropies measures. According to the sparsity of the decompositions, an optimal resolution and a reduced multi-frame are determined, defining an adapted spectrogram with variable resolution and hop size.\\
The composition of such a reduced multi-frame allows an immediate definition of a dual frame: re-synthesis techniques for this adapted analysis are easily derived by the traditional phase vocoder scheme.

\end{abstract}

\section{Introduction}
\label{sec:intro}
The quality of analysis and synthesis processes based on time-frequency transforms is highly affected by the frames used for the decomposition and the reconstruction of the signal. Traditional methods based on single frames of atomic functions have important limits: a Gabor frame imposes a fixed resolution over all the time-frequency plane, while a wavelet frame gives a strictly determined variation of the resolution: moreover, the user is frequently asked to define himself the analysis window features, which is not always a simple task even for normally experienced users.\\  
The resolution of such analysis methods is linked to the time and frequency concentration of the basic functions involved in the decomposition. Frame Theory (\cite{Gr01},\cite{Ch03},\cite{Mal99}) extends the concept of orthonormal basis in a Hilbert space $H$: in our domain, it gives a unified model for the description of decomposing systems based on atomic functions. The set $\{\phi_{\gamma}\}_{\gamma\in \Gamma}$ is a \emph{frame} for $H$ if there exist two positive non zero constants $A$ and $B$, called \emph{frame bounds}, such that for all $f\in H$, 
\begin{equation}\label{frame_complete}A\|f\|^2\leq\sum_{\gamma\in\Gamma}|\lsc f,\phi_{\gamma}\rsc|^2\leq B\|f\|^2~.\end{equation} The time-frequency concentration of an atom $\phi_{\gamma}$ in a frame can be represented through its associated Heisenberg box: it is a rectangle drawn in the time-frequency plane whose dimensions are linked respectively to the time spread of a function and to the frequency spread of its Fourier Transform. In the Short Time Fourier Transform, the boxes associated to the transpositions of the window function $g$ have fixed dimensions in every area of the time-frequency plane: the resolution is the same for all the components of the signal. In the Wavelet Transform, lower frequency components are represented with a higher time resolution, while a higher frequency resolution is given for the higher frequency ones. This limits are not motivated when analyzing a sound without an a priori knowledge of its features, as the best resolution tradeoff is neither unique nor depending only on a single variable. It is therefore useful to search for adaptive methods of sound analysis and synthesis, and for algorithms whose operations are designed to change locally according to the analyzed signal features.\\
Given $l\in\m{R}^+$, the analysis resolution can be globally modified with a scaling operation
\beq\label{scaling} g^{l}(t) = \frac{1}{\sqrt{l}}~g\bigg(\frac{t}{l}\bigg)~, \eeq
which has the effect of changing the ratio between the edges of the Heisenberg box associated to $g$ while preserving its area: this means that the global time-frequency resolution is modified by privileging concentration in one dimension to the detriment of the other. The idea which has lead to the definition of \emph{multiple Gabor frames} (\cite{Do02}) is to consider a decomposing system where all these different resolution tradeoffs coexist, providing a more detailed description of the signal. The drawback is the introduction of a high redundancy which lowers the readability of the representation: therefore methods for appropriate reductions of these multiple frames are needed, typically using sparsity criteria.\\ 
A promising approach (\cite{BF01}) takes into account \emph{Rnyi entropies}, a generalization of the Shannon entropy: given a unit-energy signal $f \in$ \ldue and a time-frequency representation $\Phi_f (u,\xi)$ of $f$ the Rnyi entropy of the representation is defined for an \emph{order} $\alpha > 0$ as follows
\beq\label{ren_ent_def}  H_{\alpha}(\Phi_f) =\frac{1}{1-\alpha}~\log_2 \iint \Phi_f^{\alpha} (u,\xi)\de u \de \xi~. \eeq
In this paper, the time-frequency representation $\Phi_f (u,\xi)$ considered is the \emph{spectrogram}, as detailed in the next section. The application to our problem is related to the concept that minimizing the complexity or information over a set of time-frequency representations of a same signal is equivalent to maximizing the concentration and peakiness of the analysis, thus selecting the best resolution tradeoff: a sparsity measure can consequently be defined through an information measure. Methods inspired by this approach have shown to give interesting results both analytically and numerically (\cite{Ja05}).\\

The proposed method of local time adaptation improves on the analysis multi-frame definition: the user can specify a finite arbitrary set of positive scaling factors $L\subset \m{R}^{+}$ corresponding to the resolutions available; then the algorithm composes different frames $\{g^l_{n,k}\}_{(n,k)\in\m{Z}^2}$ with $l\in L$ and $g^l$ as in \eqref{scaling}, and a multiple Gabor frame is obtained as the union of all the given frames. The main improvement in comparison with \cite{LT06} is that we are not obliged to keep the same hop size within the individual frames analyses, thus avoiding unnecessary short hops for larger windows: our method employs frames which share the same redundancy, so that every analysis has the same overlap, with a significant gain in computational cost.\\
The limit of our approach in comparison with \cite{Ja05} is that we apply the entropy evaluation on the whole frequency dimension, thus providing analyses which are adapted only in the time dimension. On the other hand, the reduced multi-frame obtained with our method allows a perfect reconstruction of the signal which is not provided by \cite{Ja05}: in our scheme, for any analysis segment a single original frame is retained; therefore, a re-synthesis technique can be defined as a straightforward extension of the least square error estimation from the modified STFT presented in \cite{GL84}. So our method can easily be used to provide common time-frequency processing frameworks with an adaptive analysis technique.

\section{Entropy Evaluation of a Spectrogram}
We will now describe the application of the entropy sparsity measure on the spectrogram distribution. We will focus on discretized spectrograms, as dealing with digital signal processing requires to work with sampled signals and distributions, even if for the most part the results can be extended to the continuous case.\\
A \emph{Gabor frame} is obtained by time shifting and frequency transposing a window function $g$ according to a regular lattice. Given a time step $a$ and a frequency step $b$ we write $\{u_n\}_{n\in\m{Z}} = an$ and $\{\xi_k\}_{k\in\m{Z}} = bk$; these two sequences generate the nodes of the time-frequency lattice for the frame $\{g_{n,k}\}_{(n,k)\in\m{Z}^2}$ defined as
\beq\label{gabor_frame_def} g_{n,k}(t) = g(t - u_n)e^{2\pi i\xi_k t}~;
\eeq
the nodes are the centers of the Heisenberg boxes associated to the windows in the frame. The decomposition of a function $f\in\mldue$ in a Gabor frame is simply a sampling of its STFT according to such a lattice,
\beq\label{disc_stft_def} \mathrm{S}f[n,k] = \lsc f,g_{n,k} \rsc = \int f(t)g(t-u_n)e^{-2\pi i\xi_k t}\de t~,\eeq
and the squared modulus of this decomposition is the discretized spectrogram,
\beq\label{disc_spec_def}  \mathrm{PS}_f[n,k] = |\mathrm{S}f[n,k]|^2~.\eeq  
Given a discrete spectrogram with time step $a$ and frequency step $b$ as in \eqref{disc_spec_def}, we look for an evaluation of its entropy over a certain rectangle of the time-frequency plane $[t_1,t_2]\times[\nu_1,\nu_2] \subseteq \m{R}^2$. The rectangle identifies a sequence of points $G \subseteq \m{Z}^2$ where $G = \{(n,k)\in\m{Z}^2: t_1 \leq na \leq t_2,~ \nu_1 \leq kb \leq \nu_2\}$. Through an appropriate normalization we obtain the sequence
\beq\label{local_spec} \mathrm{PS}_f^G[n,k] = \frac{\mathrm{PS}_f[n,k]}{\sum_{[n',k']\in G} \mathrm{PS}_f[n',k']} ~,\eeq
with $[n,k]\in G$, which can be seen as a discrete probability density. As a discretization of the original continuous spectrogram, every sample in $\mathrm{PS}_f^G$ is related to a time-frequency region of area $ab$; we thus obtain the Rnyi entropy measure for \eqref{local_spec} directly from \eqref{ren_ent_def},
\beq\label{ren_ent_disc} H_{\alpha}(\mathrm{PS}_f^G ) =  \frac{1}{1-\alpha}\log_2 \sum_{[n,k]\in G}(\mathrm{PS}_f^G[n,k])^{\alpha}  + \log_2(ab)~.\eeq
General properties of Rnyi entropies can be found in \cite{Re61}, \cite{BS93} and \cite{Zy04}; we recall in particular those which have a closer relation with our problem. It is easy to show that for every finite discrete probability density $P$ the entropy $H_{\alpha}(P)$ tends to coincide with the Shannon entropy of $P$ as the order $\alpha$ tends to one. Moreover, $H_{\alpha}(P)$ is a non increasing function of $\alpha$, so
\beq\label{ren_ent_nonincalpha} \alpha_1 < \alpha_2 \Rightarrow H_{\alpha_1}(P)\geq H_{\alpha_2}(P)~.\eeq
As we are working with finite discrete densities we can also consider the case $\alpha = 0$ which is simply the logarithm of the number of elements in $P$; as a consequence $H_0(P) \geq H_{\alpha}(P)$ for every admissible order $\alpha$.\\
A third basic fact is that for every order $\alpha$ the Rnyi entropy $H_{\alpha}$ is maximum when $P$ is uniformly distributed, while it is minimum and equal to zero when $P$ has a single non-zero value. Given a generic $P$ and its entropy $H_{\alpha}(P)$ for a certain order $\alpha$, we have that for any $\beta \geq \alpha$
\beq\label{ren_ent_relalpha} \frac{\alpha - 1}{\alpha} H_{\alpha}(P) \leq \frac{\beta - 1}{\beta} H_{\beta}(P)~. \eeq	
 All of these results give useful informations on the values of different measures on a single density $P$ as in \eqref{ren_ent_disc}, while the relations between the entropies of two different densities $P$ and $Q$ are in general hard to determine analytically; in our problem,  $P$ and $Q$ are two spectrograms of a same signal in a same time-frequency area, based on two window functions with different scaling as in \eqref{scaling}.\\
 When the spectrogram of a signal does not depend on time it is easier to find such a relation, and it turns out to be the one expected: let $\mathrm{PS}_s^G$ be the sampled spectrogram of a sinusoid $s$ over the region $G$ with a window function $h$ of compact support; then $\mathrm{PS}_s^G$ is simply a translation in the frequency domain of $\ca{h}$, the Fourier transform of the window, and it is therefore time-independent. We choose a bounded set $L$ of admissible scaling factors, so that the discretized support of the scaled windows $h^l$ still remains inside $G$ for any $l\in L$. It is not hard to prove that the entropy of a spectrogram taken with such a scaled version of $h$ is given by
 \beq\label{ren_ent_scaledspec} H_{\alpha}(\mathrm{PS}_{sl}^G ) = H_{\alpha}(\mathrm{PS}_s^G ) - \log_2 l~.\eeq
 The sparsity measure we are using looks for the window which minimizes the entropy measure: we deduce from \eqref{ren_ent_scaledspec} that it is the one obtained with the largest scaling factor available, so with the largest time-support. This is coherent with our expectation as stationary signals, such as sinusoids, are best analyzed with a high frequency resolution, because time-independency allows a small time resolution. Moreover, this is true for any order $\alpha$ used for the entropy calculus.\\
Symmetric considerations apply whenever the spectrogram of a signal does not depend on frequency, as for impulses.\\

A last remark regards the dependency of \eqref{ren_ent_disc} on the time and frequency step $a$ and $b$ used for the discretization of the spectrogram. When considering signals as finite vectors, a signal and its Fourier Transform have the same length. Therefore in the STFT the window length determines the number frequency points, while the sampling rate sets frequency values: the definition of $b$ is thus implicit in the window choice. Actually, the FFT algorithm allows to ask a number of frequency points larger than the signal length: further frequency values are obtained as an interpolation between the original ones by properly adding zero values to the signal. If the sampling rate is fix, such a procedure establishes smaller $b$ as a consequence of a larger number of frequency points. We have numerically verified that such a variation of $b$ has no impact on the entropy calculus, so that the FFT size can be set according to implementation needs.\\
Regarding the time step $a$, we are working on the analytical demonstration of a largely verified evidence: as long as the decomposing system is a frame the entropy measure is invariant to redundancy variation, so the choice of $a$ can be ruled by considerations on the invertibility of the decomposing frame without losing coherence between the information measure of the different analyses. This is a key point, as it states that the sparsity measure obtained allows a total independence between the hop sizes of the different analyses: with the implementation of proper structures to handle multi-hop STFTs we have obtained a more efficient algorithm in comparison with the ones imposing a fixed hop size, as \cite{LT06}.

\section{Algorithm and examples}

We now summarize the main operations of  the algorithm we have developed providing examples of its application. For spectrograms calculation we have used a \emph{Hanning window}
\beq\label{han_def} h(t) = \cos^2(\pi t)\chi_{[-\frac{1}{2},\frac{1}{2}]}~,\eeq
with $\chi$ the indicator function of the specified interval, but it is obviously possible to generalize the results thus obtained to the entire class of compactly supported window functions. We create a multiple Gabor frame as in \eqref{gabor_frame_def} using as mother functions some scaled version of $h$, obtained as in \eqref{scaling} with a finite set of positive real scaling factors $L$.\\
Different spectrograms of segments of the signal are calculated with each one of the above frames: the length of the analysis segment and the overlap between two consecutive segments are given as parameters.\\

\begin{figure}[ht]
\centerline{\includegraphics[height=5.4cm]{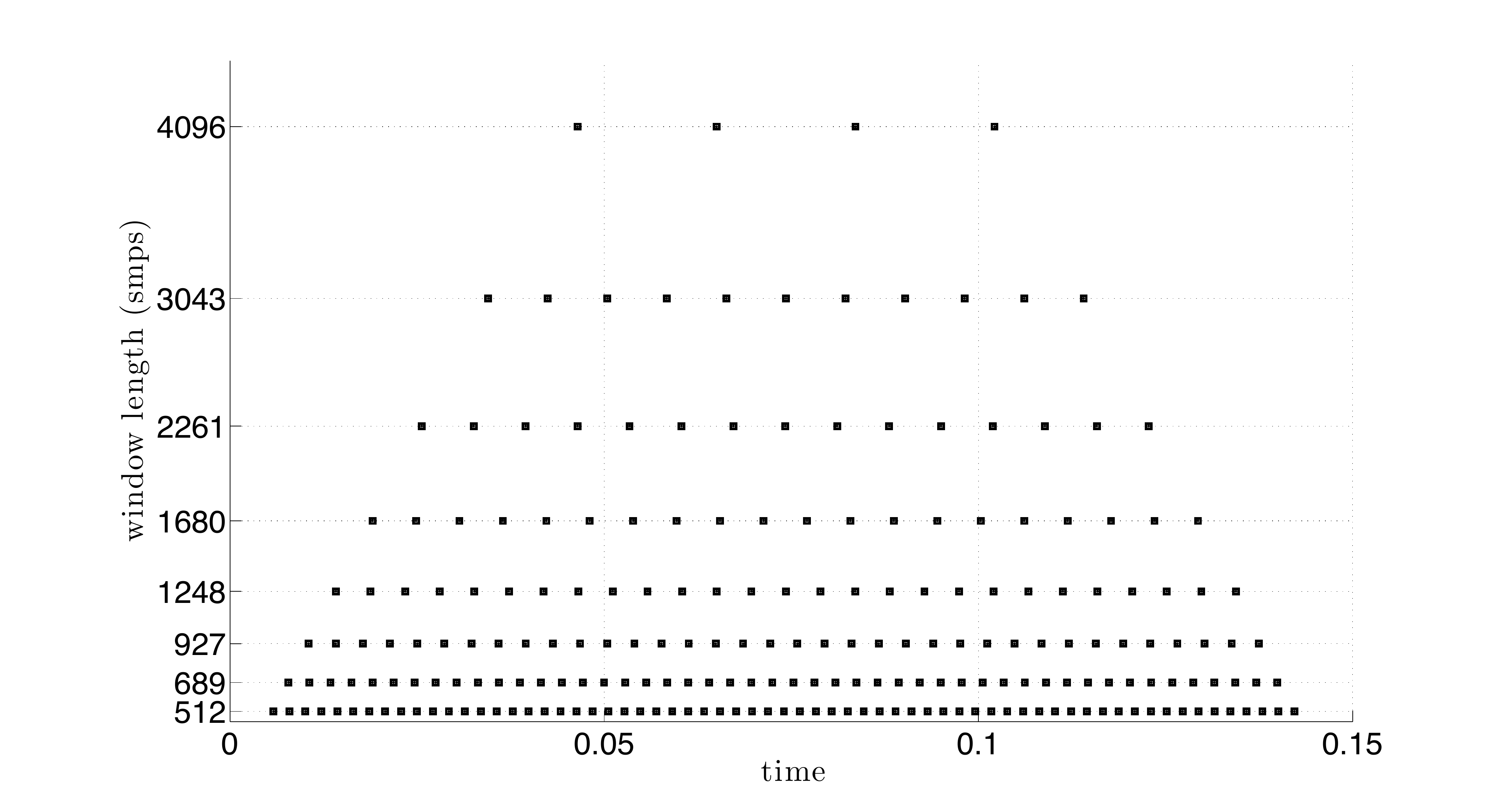}}
\caption{\label{fig1} {\it An analysis segment: time locations of the Heisenberg boxes associated to the multi-frame used in our algorithm.}}
\end{figure}

The different frames composing the multi-frame have the same frequency step $b$ but different time steps $\{a_l: l\in L\}$: the smallest and largest window sizes are given as parameters together with $|L|$, the number of different windows composing the multi-frame, and the global overlap needed for the analyses. The algorithm fixes the intermediate sizes so that for each signal segment the different frames have the same overlap between consecutive windows, and so the same redundancy. This generates an irregular time disposition of the multi-frame elements in each signal segment, as illustrated in figure \ref{fig1}. Such a disposition causes a different influence of the boundary parts of the signal on the different frames analyses: the beginning and the end of the signal segment have a higher energy when windowed in the smaller frames. This is avoided with a preliminary weighting: the beginning and the end of each signal segment are windowed respectively with the first and second half of the largest analysis window. Such a weighting does not concern the decomposition for re-synthesis purpose, but only the analyses used for entropy evaluations. 
For each signal segment we calculate the entropy of every spectrogram as in \eqref{ren_ent_disc}, where $G$ is the rectangle with the time segment analyzed as horizontal dimension and the whole frequency lattice as vertical. The sparsest local analysis is defined to be the one with minimum Rnyi entropy: the best window is thus defined consequently. Adaptation is obtained over the time dimension as for every signal segment the selected analysis involve the whole frequency dimension. An interpolation is performed over the overlapping zones to avoid abrupt discontinuities in the tradeoff of the resolutions.\\
The time adapted analysis of the global signal is finally realized by opportunely assembling the slices of local sparsest analyses obtained with the selected windows.\\

\begin{figure*}[ht]
\center
\includegraphics[height=9.6cm]{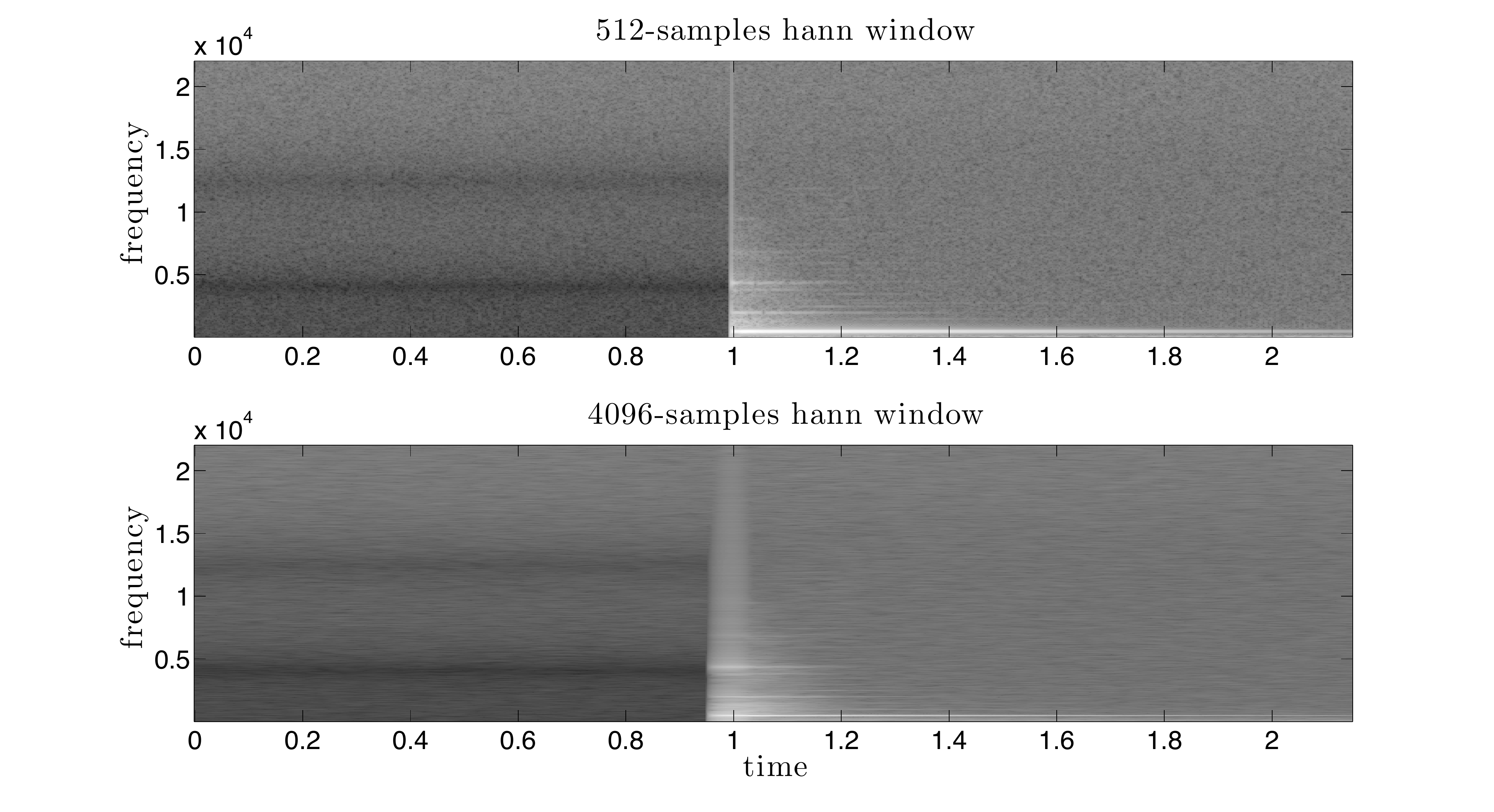}
\caption{\label{fig2}{\it Two different spectrograms of a B4 note played by a marimba, with Hanning windows of sizes 512 (top) and 4096 (bottom) samples.}}
\end{figure*}

\begin{figure*}[ht]
\center
\includegraphics[height=9.6cm]{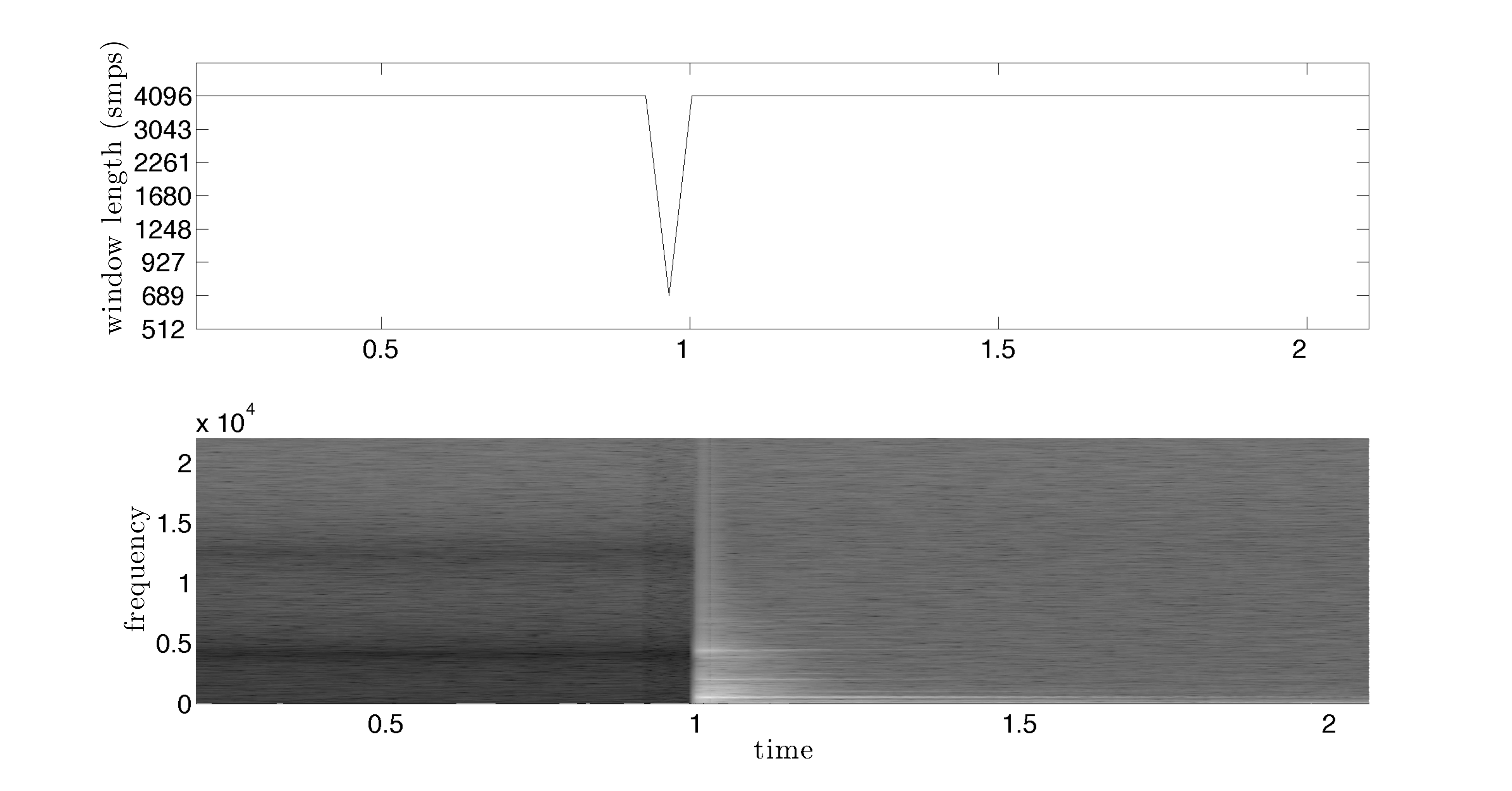}
\caption{\label{fig3}{\it Example of an adaptive analysis performed by our algorithm with eight Hanning windows of different sizes from 512 to 4096 samples, on a B4 note played by a marimba sampled at 44.1kHz: on top, the best window chosen as a function of time; at the bottom, the adaptive spectrogram. The entropy order is $\alpha = 0.7$ and each analysis segment contains four frames of the largest window analysis with a two-frames overlap between consequent segments.}}
\end{figure*}

In figure \ref{fig3} we give a first example of an adaptive analysis performed by our algorithm with eight Hanning windows of different sizes on a real instrumental sound, a B4 note played by a marimba: this sound combines the need for a good time resolution at the moment of the percussion, with that of a good frequency resolution on the harmonic resonance of the instrument. This is fully provided by the algorithm, as shown in the adapted spectrogram at the bottom of figure \ref{fig3}. Moreover, we see that the pre-echo of the analysis at the bottom of figure \ref{fig2} is completely removed in the adapted spectrogram.\\

\begin{figure*}[ht]
\center
\includegraphics[height=9.6cm]{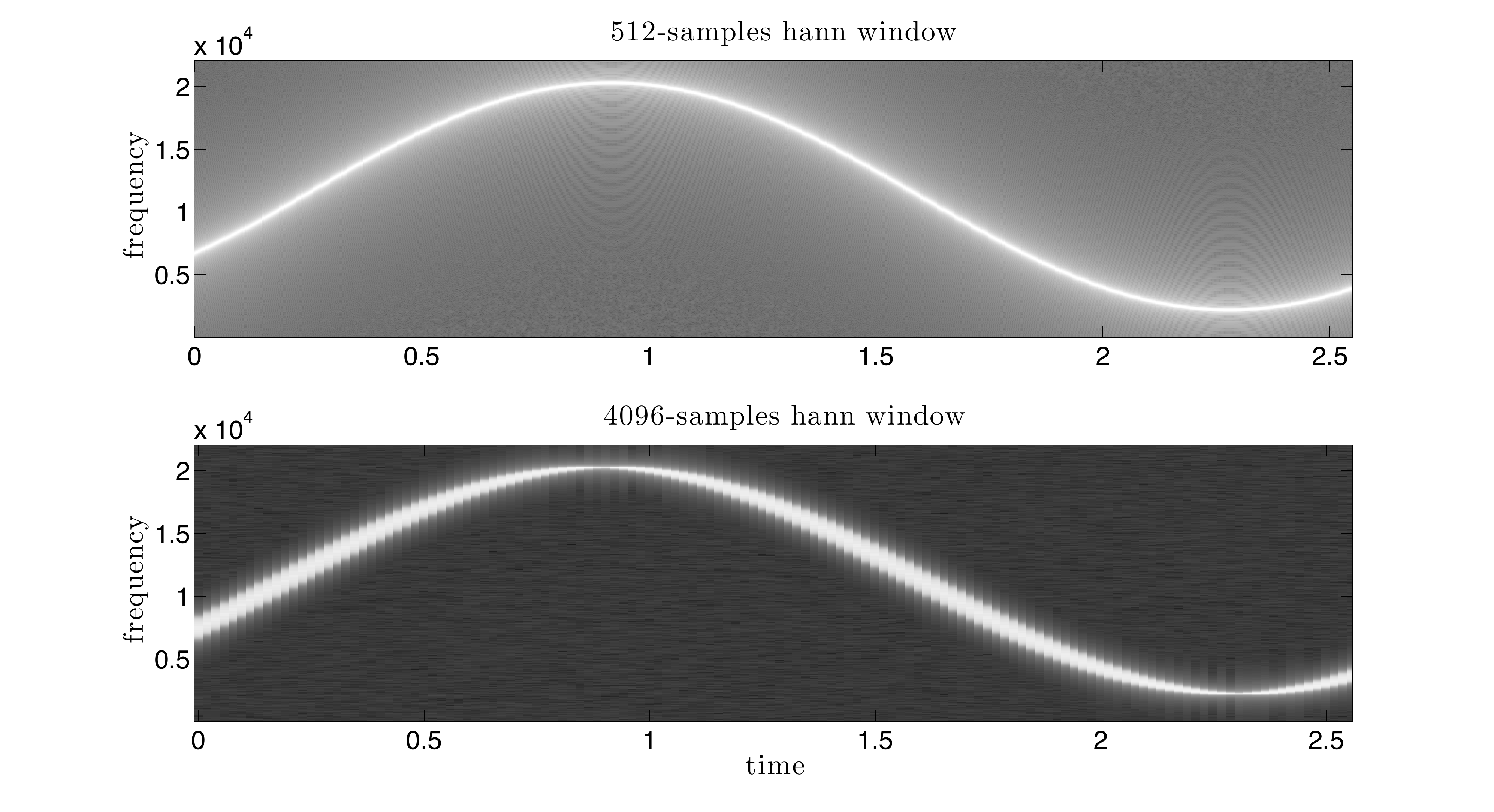}
\caption{\label{fig4}{\it Two different spectrograms of a sinusoid with sinusoidal frequency modulation, with Hanning windows of sizes 512 (top) and 4096 (bottom) samples. }}
\end{figure*}

\begin{figure*}[ht]
\center
\includegraphics[height=9.6cm]{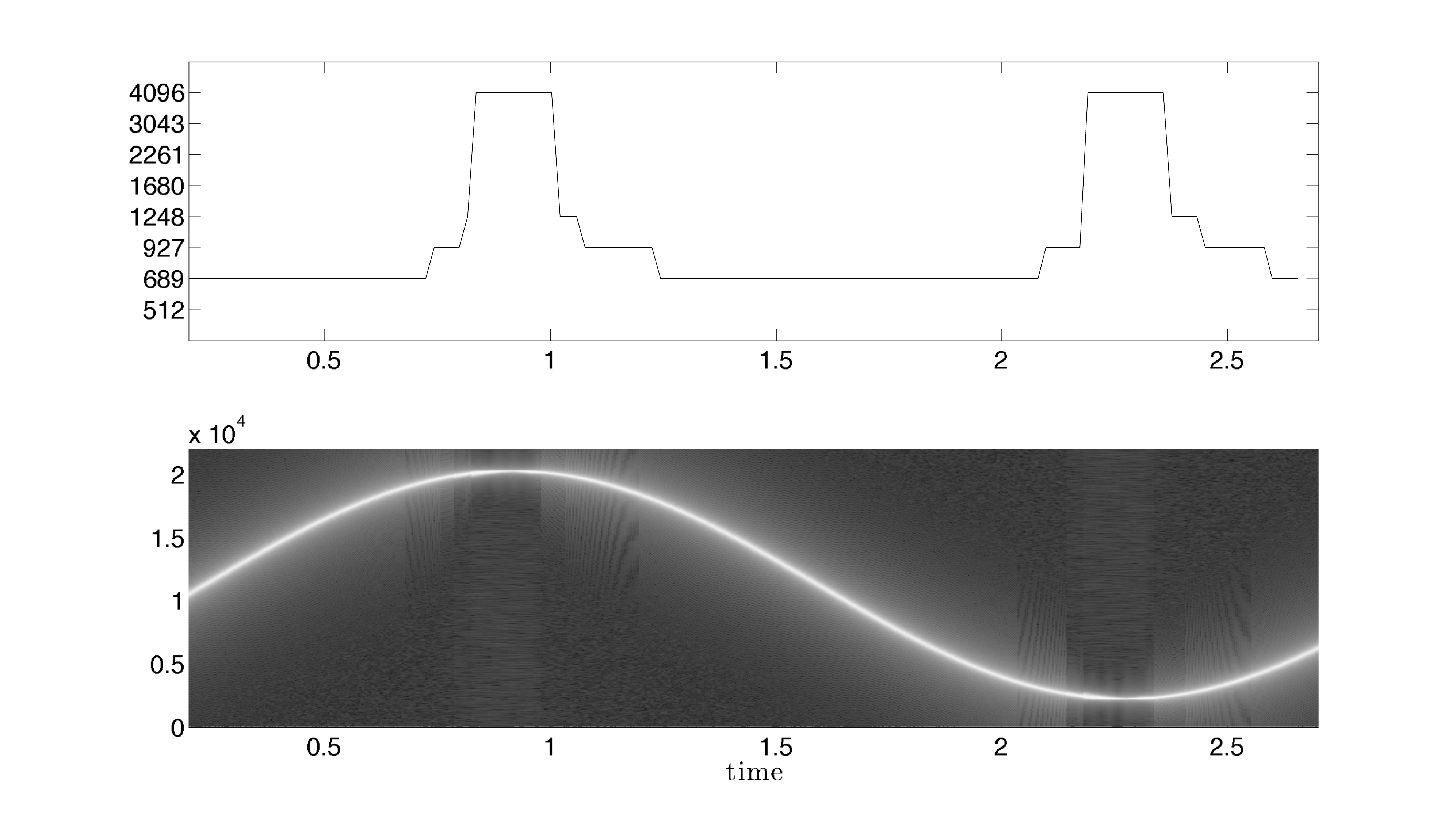}
\caption{\label{fig5}{\it Example of an adaptive analysis performed by our algorithm with eight Hanning windows of different sizes from 512 to 4096 samples, on a sinusoid with sinusoidal frequency modulation synthesized at 44.1 kHz: on top, the best window chosen as a function of time; at the bottom, the adaptive spectrogram. The entropy order is $\alpha = 0.7$ and each analysis segment contains four frames of the largest window analysis with a three-frames overlap between consequent segments.}}
\end{figure*}

In figure \ref{fig5} we give a second example with a synthetic sound, a sinusoid with sinusoidal frequency modulation: as figure \ref{fig4} shows, a small window is best adapted where the frequency variation is fast compared to the window length; on the other hand,  the largest window is better where the signal is almost stationary.


The re-synthesis method introduced in \cite{GL84} gives a perfect reconstruction of the signal as a weighted expansion of the coefficients of its STFT in the original analysis frame. Let $S_f[n,k]$ be the STFT of a signal $f$ with window function $h$ and time step $a$; fixing $n$, through an iFFT we have a windowed segment of $f$
\beq\label{windowed_sig} f_h(n,l) = h(na-l)f(l)~,\eeq
whose time location depends on $n$. An immediate perfect reconstruction of $f$ is given by
\beq\label{perf_rec} f(l) = \frac{\sum_{n = -\infty}^{+\infty} h(na - l)f_h(n,l)}{\sum_{n = -\infty}^{+\infty}h^2(na - l)}~.\eeq
We extend the same technique using a variable window $h$ and time step $a$ according to the composition of the reduced multi-frame, obtaining a perfect reconstruction as well. The interest of \eqref{perf_rec} is that the given distribution needs not to be the STFT of a signal: for example, a transformation $S^*[n,k]$ of the STFT of a signal could be considered. In this case, \eqref{perf_rec} gives the signal whose STFT has minimal least squares error with $S^*[n,k]$.\\
The theoretical existence and the mathematical definition of the canonical dual frame for reduced multi-frames like the one we employ has recently been provided in \cite{JBD09}: the analysis and re-synthesis framework is thus entirely defined within the Gabor theory, but no automatic adaptation is employed. We are at present working on the interesting analogies between the two approaches, to establish a unified interpretation and develop further extensions.
%
\section{Conclusions}
We have presented an algorithm for time-adaptation of the spectrogram resolution, which can be easily integrated in existent framework for analysis, transformation and re-synthesis of an audio signal: the adaptation is locally obtained through an entropy minimization within a finite set of resolutions,  which can be defined by the user or left as default. The user can also specify the time duration and overlap of the analysis segments where entropy minimization is performed, to privilege more or less discontinuous adapted analyses.\\
Future improvements of this method will concern the spectrogram adaptation in both time and frequency dimensions: this will provide a decomposition of the signal in several layers of analysis frames, thus requiring an extension of the proposed technique for re-synthesis.

\nocite{*}
\bibliographystyle{IEEEbib}
\bibliography{lrrr_template} 

\begin{thebibliography}{10}

\bibitem{Gr01}
K.~Gr\"ochenig, Ed.,
\newblock {\em Foundations of Time-Frequency Analysis},
\newblock Birkh\"auser, Boston, Massachusetts, USA, 2001.

\bibitem{Ch03}
O.~Christensen, Ed.,
\newblock {\em An Introduction To Frames And Riesz Bases},
\newblock Birkh\"auser, Boston, Massachusetts, USA, 2003.

\bibitem{Mal99}
S.~Mallat, Ed.,
\newblock {\em A wavelet tour on signal processing},
\newblock Academic Press, San Diego, California, USA, 1999.

\bibitem{Do02}
M.~D\"orfler,
\newblock {\em Gabor Analysis for a Class of Signals called Music},
\newblock Ph.D. thesis, Institut fr Mathematik der Universitt Wien, 2002.

\bibitem{BF01}
R.G. Baraniuk P. Flandrin A.J.E.M. Janssen~O.J.J. Michel,
\newblock ``Measuring {T}ime-{F}requency {I}nformation {C}ontent {U}sing the
  {R}\'enyi {E}ntropies,''
\newblock {\em IEEE Trans. Info. Theory}, vol. 47, no. 4, pp. 1391--1409, May
  2001.

\bibitem{Ja05}
F.~Jaillet,
\newblock {\em Reprsentation et traitement temps-frquence des signaux
  audionumriques pour des applications de design sonore},
\newblock Ph.D. thesis, Universit de la Mditerrane - Aix-Marseille II, 2005.

\bibitem{LT06}
J.Todd A.Lukin,
\newblock ``{Adaptive Time-Frequency Resolution for Analysis and Processing of
  Audio},''
\newblock 120th Audio Engineering Society Convention, Paris, France, May 2006.
  \href{PDF}{http://graphics.cs.msu.ru/en/publications/text/LukinTodd.pdf}.

\bibitem{GL84}
D.W. Griffin~J.S. Lim,
\newblock ``Signal {E}stimation from {M}odified {S}hort-{T}ime {F}ourier
  {T}ransform,''
\newblock {\em IEEE Trans. Acoust. Speech Signal Process.}, vol. 32, no. 2, pp.
  236--242, Apr. 1984.

\bibitem{Re61}
A.~Rnyi,
\newblock ``On {M}easures of {E}ntropy and {I}nformation,''
\newblock in {\em Proc. Fourth Berkeley Symp. on Math. Statist. and Prob.},
  Berkeley, California, June 20-30, 1961, pp. 547--561.

\bibitem{BS93}
F.~Schl\"ogl C.~Beck, Ed.,
\newblock {\em Thermodynamics of chaotic systems},
\newblock Cambridge University Press, Cambridge, Massachusetts, USA, 1993.

\bibitem{Zy04}
K.~Zyczkowski,
\newblock ``Rnyi {E}xtrapolation of {S}hannon {E}ntropy,''
\newblock {\em Open Systems \& Information Dynamics}, vol. 10, no. 3, pp.
  297--310, Sept. 2003.

\bibitem{JBD09}
F.~Jaillet P. Balazs M. D\"orfler~N. Engelputzeder,
\newblock ``Nonstationary {G}abor {F}rames,''
\newblock in {\em Proc. of SAMPTA'09}, Marseille, France, May 18-22, 2009.

\end{thebibliography}
 
\end{document}